\documentclass[11pt]{article}
\usepackage{amsmath,amssymb,amsbsy,amstext,amsthm,booktabs,simplewick,exscale,relsize,slashed,graphicx,amsfonts,upgreek,placeins}
\usepackage{multirow,dcolumn,bm,enumerate,url,hyperref,cite}
\usepackage{color}
\usepackage[a4paper, total={6.5in, 9.5in}]{geometry}
\usepackage[utf8]{inputenc}
\usepackage{graphicx}
\usepackage{caption}
\usepackage[T1]{fontenc}
\usepackage{multirow}
\usepackage{mathtools}
\usepackage{authblk}
\usepackage{csquotes}
\usepackage[english]{babel}
\usepackage{subcaption}
\usepackage{amsmath}
\usepackage{amssymb}
\usepackage{multirow}
\usepackage{float}

\DeclarePairedDelimiter{\evdel}{\langle}{\rangle}
\newcommand{\ev}{\evdel}

\title{\textbf{Terrestrial and Martian Heat Flow Limits on Dark Matter}}
\author[1,2]{Joseph Bramante}
\author[1]{Andrew Buchanan}
\author[1]{Alan Goodman}
\author[1]{Eesha Lodhi}

\affil[1]{The Arthur B. McDonald Canadian Astroparticle Physics Research Institute, \protect\\ Department of Physics, Engineering Physics, and Astronomy, \protect\\ Queen's University, Kingston, Ontario, K7L 2S8, Canada}
\affil[2]{Perimeter Institute for Theoretical Physics, Waterloo, Ontario, N2L 2Y5, Canada}
\date{\today}

\begin{document}

\maketitle

\begin{abstract}
If dark matter is efficiently captured by a planet, energy released in its annihilation can exceed that planet's total heat output. Building on prior work, we treat Earth's composition and dark matter capture in detail and present improved limits on dark matter-nucleon scattering cross sections for dark matter masses ranging from 0.1 to $10^{10}$ GeV. We also extend Earth limits by applying the same treatment to Mars. The scope of dark matter models considered is expanded to include spin-dependent nuclear interactions including isospin-independent, proton only, and neutron only interactions. We find that Earth and Mars heating bounds are alleviated for dark matter s-wave self-annihilation cross sections $\lesssim 10^{-44}~{\rm cm^2}$.
\end{abstract}

\section{Introduction}
Despite a preponderance of evidence for the existence of dark matter (DM), its mass and the nature of its non-gravitational interactions remain unknown. It may be the case that DM couples to Standard Model fields through a non-gravitational force. Here we will study DM's interactions with itself and with nucleons in the Earth and Mars, and by precisely modeling each, derive extended planetary heating bounds, first obtained in Reference \cite{Mack2007}.

Many underground experiments have searched for DM that scatters with nucleons through either predominantly spin-dependent or spin-independent interactions, including DEAP \cite{Ajaj:2019imk}, PICO \cite{PICO1,PICO2}, LUX \cite{LUX}, PandaX \cite{PandaX}, and XENON1T \cite{XENON1T}. These experiments place detectors deep underground to reduce backgrounds and gain in sensitivity to weakly interacting DM. However, such experiments situated kilometers underground are less sensitive to more strongy-interacting DM, which has reduced kinetic energy after repeated scattering against the Earth's crust during its voyage underground. On the other hand, near-surface direct detection experiments \cite{Overburden} are more sensitive to strongly-interacting DM, along with repurposed high-altitude detectors like the XQC rocket \cite{XQC}. Constraints on strongly-interacting DM from these experiments are supplemented by numerous astrophysical bounds, including analyses of the cosmic microwave background \cite{CMB1, CMB2}, interstellar gas cooling \cite{GasClouds1, GasClouds2}, neutrinos from DM annihilation \cite{Silk:1985ax,Freese:1985qw,Gould:1987ir,Gould:1991hx,Baum:2016oow}, and a conspicuous lack of tracks in ancient mica \cite{mica1,mica2,mica3}. This article will reinforce and extend existing constraints on DM's spin-independent interactions. Additionally, there are a number of simple models for which DM's non-relativistic interactions with nuclei depend sensitively on the spin of the nucleus \cite{Goodman:1984dc,Agrawal:2010fh}. For spin-dependent models, this work places bounds on some unexcluded regions of DM parameter space; to achieve this we incorporate the terrestrial distribution of nuclei with a substantial nuclear spin.

Strongly-interacting DM can appreciably raise the temperature of the Earth and Mars through capture and subsequent annihilation. Simulations indicate that galaxies like the Milky Way exist within a virialized and spherical DM halo \cite{Primack2012, NFW2012}. From this it follows that so long as DM is a light particle (in this case lighter than a small asteroid), there will be a constant flux of DM incident upon the Earth as it orbits the Galaxy. If DM interacts sufficiently strongly with nucleons, it will scatter off elements within these planets, and if it scatters enough times, its velocity will decrease below the given planet's escape velocity, which is approximately 11 km$/$s for Earth, and 5 km$/$s for Mars. If DM is sufficiently slowed, it will stay bound to the planet and can annihilate with other similarly captured DM. Such annihilations may result in a sizable heat output. Because we know from direct measurements that no more than $\sim$44 TW of energy is emitted by the Earth\cite{williams1974, lister1990, JAUPART2015223, Davies1980, Davies1980_2, sclater1980, pollack1993, davies2010}, we are able to restrict DM candidates by requiring that they not yield more heat flow from Earth than is observed. While the origin of heat emitted by Earth is still under investigation, it is expected that around half of the observed heat emission is radiogenic \cite{davies2010}. Because in-situ heat measurements are not available for Mars and other planets, it is not yet possible to set a constraint using a planet other than the Earth. However, the InSight mission to Mars may soon report the first direct Martian heat flow measurements \cite{2013LPI....44.1915B}. Therefore, we have also studied what constraint could be placed by observing no more than the expected $\sim$3.5 TW of radiogenic energy generated within Mars \cite{parro2017}. 

The most stringent limit on DM's nucleon interactions is obtained from planetary heating when the DM annihilation rate equals the DM capture rate. This occurs for DM with a sufficiently large self-annihilation cross section. This scenario is referred to as ``total annihilation'' \cite{Mack2007}. Of course, a smaller DM annihilation cross section will result in less annihilation. This ``partial annihilation'' scenario is addressed in this article and we find that the heating bound effectively vanishes for $\sigma_{\bar \chi \chi } \lesssim 10^{-44}~{\rm cm^2}, 10^{-30}~{\rm cm^2},$ and $10^{-38}~{\rm cm^2}$, for s-wave, p-wave, and impeded dark matter annihilation respectively. Intriguingly, these cross sections are around the canonical ``weak'' scale annihilation cross section $\sigma_{\bar \chi \chi }^{weak} \approx 10^{-36} ~{\rm cm^2}$ \cite{Bertone:2004pz}. Therefore, for DM with an approximately weak scale annihilation cross section, planetary heating bounds can have a non-trivial dependence on DM's self-annihilation cross section.

This work primarily considers three DM parameters: (1) the DM mass $m_\chi$, (2) the DM per-nucleon scattering cross section $\sigma_{\chi N}$, which determines the frequency with which the DM particle will interact with terrestrial and Martian elements, and (3) DM's self-annihilation cross section $\sigma_{\bar{\chi} \chi}$, which determines the extent to which captured DM will result in increased heat flow out of the Earth and Mars. In Section \ref{sec:cap} we treat the planetary capture of strongly-interacting DM, incorporating a three zone nuclear abundance model of the Earth's core, mantle, and crust, and a two zone nuclear abundance model of Mars' core and mantle. Section \ref{sec:lims} details limits on DM from planetary heating, including both total and partial DM self-annihilation in the Earth. In Section \ref{sec:conc} we conclude. Throughout this work, we have used natural units with $\hbar = k_B = c = 1$.

\section{Dark Matter Capture}
\label{sec:cap}

As DM passes through a planet, it may scatter off its constituent particles, and will slow slightly with each scatter. For certain DM masses and cross sections, this scattering results in DM slowing below the planet's escape velocity, at which point it is gravitationally bound to the planet. 

To simplify our DM capture computations, we ignore Earth and Mars' gravitational effects on the DM's trajectory and velocity. This is a reasonable approximation, given that the speed of a typical DM particle is $\sim$300 $~{\rm km/s}$, compared to the $\sim$10$~{\rm km/s}$ escape speeds in question. We also ignore the directional changes in the DM's trajectory induced by scattering. These are both conservative approximations, as both a random walk from scattering and a gravitationally curved trajectory lead the DM through more material than a straight trajectory.

For convenience, we define
\begin{equation}
\label{betaDef}
    \beta^\pm_j=4\frac{m_jm_\chi}{(m_j\pm m_\chi)^2},
\end{equation}
\noindent where $m_\chi$ is the mass of the DM particle and $m_j$ is the mass of the terrestrial constituent $j$ off of which it is scattering. The DM particle's kinetic energy after a single scatter, $E_f$, can be defined as a function of its initial kinetic energy, $E_i$, as
\begin{equation}
\label{EfEi}
    E_f=(1-z\beta^+_j) E_i,
\end{equation}
\noindent where $z \subset [0,1]$ is a kinematic factor parameterizing the scattering angle of the DM-nuclear interaction \cite{Bramante2018}. On average $\langle z \rangle \approx \frac{1}{2}$, and we set it to this value in our computations. Making the substitution that kinetic energy is proportional to velocity ($v$) squared, and now considering $\tau_j$ scatters off of each element $j$, the expression for DM's final velocity is

\begin{equation}
    v_f= v_i\prod_{j} (1-z\beta^+_j)^{{\tau_j}/2}.
    \label{eq:vf}
\end{equation}
When $v_f < 11.2 $ km s$^{-1}$ or $5.0 $ km s$^{-1}$ for Earth or Mars respectively, the DM is gravitationally captured, as it is no longer traveling fast enough to escape the planet's gravity. However, see later sections for discussion of DM evaporation.

\begin{figure}[ht]
    \begin{center}
        \includegraphics[width=0.4\linewidth]{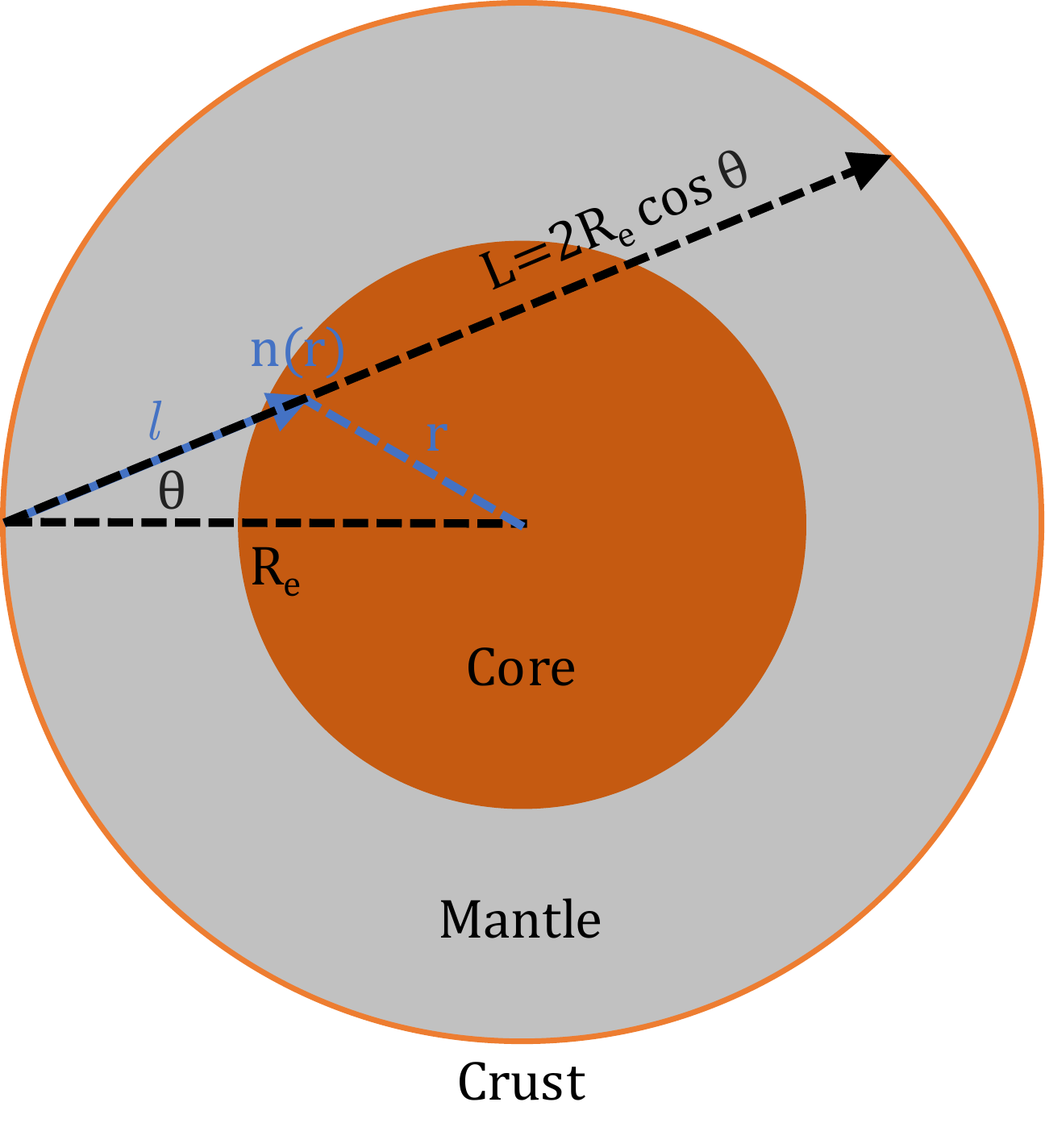}
        \captionsetup{justification=centering}
        \caption{A schematic representation of the Earth with geometric variables as labelled.} 
        \label{fig:Earthmodel}
    \end{center}
\end{figure}

To find an expression for $\tau_j$, we must first find the \textit{average} number of scatters off of each element. When travelling a distance $L$ through a medium of constant density, we define this as $\ev{\tau_j}=n_j\sigma_{\chi j}L$ \cite{Bramante2018}, where $\sigma_{\chi j}$ is the nuclear cross section of the DM with element $j$. In a medium with non-constant elemental density like the Earth and Mars, we instead must define $\ev{\tau_j}$ as a function of $\theta$, the angle between the DM particle's trajectory and the vector normal to the planet's surface. To find the mean number of scatters, we integrate along the DM's path ($l$) from its entry point to a distance of 2$R_e\cos(\theta)$, at which point it will exit the planet. A schematic diagram of the trajectory through the Earth is given in Figure \ref{fig:Earthmodel}. Explicitly, the expectation value for the number of scatters is given by the integral

\begin{equation}
\label{taubar}
    \ev{\tau_j}(\theta)= \sigma_{\chi j} \int_{0}^{2R_e\cos{\theta}} n_j(r)\cdot dl.
\end{equation}

Because number density, $n_j$, is a function of radius from the planet's center, the substitution $r=\sqrt{l^2+R_e^2-2 l R_e \cos{\theta}}$ must be made. Scattering events are independent and discrete, so $\tau_j(\theta)$ actually follows a Poisson distribution with a mean of $\ev{\tau_j}$. In computations we required the number of scatters off of element $j$ to follow the probability distribution
\begin{equation}
\label{poisson}
    p(\tau_j;\ev{\tau_j})=\frac{e^{-\ev{\tau_j}}\ev{\tau_j}^{\tau_j}}{\tau_j!}.
\end{equation}

It is also important to distinguish between DM-nucleus cross section ($\sigma_{\chi j}$) and the DM-nucleon cross section ($\sigma_{\chi N}$). We follow the conventions of \cite{Kurylov2004,Bramante2018,Mack2007}. For a spin-independent cross section we used the conversion

\begin{equation}
    \sigma_{\chi j}^{(SI)}=A_j^2\left(\frac{\mu(m_j)}{\mu(m_N)}\right)^2 \sigma_{\chi N}^{(SI)} \ \ \propto \ \ A_j^4\sigma_{\chi N}^{SI},
    \label{eq:si}
\end{equation}

\noindent where $m_N$ is the mass of a nucleon ($\sim$1 GeV), $\mu(m_j \textrm{ or } m_N)$ is the reduced mass of the DM particle and atom $j$ or a single nucleon respectively, and $A_j$ is the number of nucleons in atom $j$. In the spin-dependent case,

\begin{equation}
\label{sigmaSD}
    \sigma_{\chi j}^{(SD)}=\left(\frac{\mu(m_j)}{\mu(m_N)}\right)^2 \frac{4(J_j+1)}{3J_j} \left[a_p\ev{S_p}_j + a_n\ev{S_n}_j\right]^2 \sigma_{\chi N}^{(SD)} \ \ \propto \ \ A_j^2\sigma_{\chi N}^{(SD)}.
\end{equation}
Here, we define $J_j$ as nuclear spin of atom $j$, $\ev{S_p}_j$ and $\ev{S_n}_j$ as its average proton and neutron spins, and $a_p$ and $a_n$ as proton and neutron coupling constants. In this work, we consider three cases: (1) isospin-independent scattering ($a_p=a_n=1$), (2) proton-only scattering ($a_p=1$, $a_n=0$), and (3) neutron only scattering ($a_p=0$, $a_n=1$).

Finally, we note that in a number of recent publications \cite{Bramante2018,mica2,Digman:2019wdm}, it has been pointed out that the spin-independent DM-nucleon scattering cross section as presented in Eq.~\eqref{eq:si}, reaches a theoretical transition point at $\sigma_{\chi N} \approx 10^{-26} ~{\rm cm^2}$. At larger cross sections, the implied DM-nucleus cross section is larger than the physical area of nuclei, implying either long-range forces ($e.g.$ mediation by light dark photons \cite{GasClouds2,Digman:2019wdm}) or composite dark matter \cite{Coskuner:2018are}. 

\section{Elemental Makeup}
\label{sec:Earthcomp}

The Earth's material composition is divided into three parts: the crust, the mantle, and the core. Taking the Earth's center as $r=0$, the crust begins at the Earth's surface at $r=R_e=6371$ km and ends at $r=6346$ km \cite{clarke1924composition}. From that radius until $r=3480$ km is the mantle, \cite{WANG2018460}
and the remainder of the Earth is the core in our model \cite{Morgan6973}.
The relative sizes of these regions are shown in Figure \ref{fig:Earthmodel}, and the material composition of these regions is given below in Table \ref{tab:SIcomp}.

Also shown in Table \ref{tab:SIcomp} are the compositions of Mars' mantle and core. We assume Mars' crust to have a thickness of 50 km \cite{JOHNSTON1977}. Its composition is taken as the same as that of the mantle for calculating DM drift times in section \ref{driftTime}, and is conservatively omitted for capture and annihilation computations in sections \ref{perfAnn} and \ref{impann}. We give the core and mantle thicknesses of 2000 and 1340 km respectively, for a total Martian radius of $R_m=3390$ km. All radial thickness values given here have been chosen among values presented in the above references, to minimize DM capture on Earth and Mars.

\begin{table}[ht]
\begin{center}
\captionsetup{justification=centering}
\caption{Rounded weight percentages of elements of interest in the crust, mantle, and core \cite{clarke1924composition,Morgan6973,WANG2018460,McDonough2003,johnston1974}.} 
\label{tab:SIcomp}
\resizebox{\textwidth}{!}{
\begin{tabular}{c c c c c c c c c c c c c c}\hline
& \textbf{$^{16}$O} & \textbf{$^{28}$Si} & \textbf{$^{27}$Al} & \textbf{$^{56}$Fe} & \textbf{$^{40}$Ca} & \textbf{$^{23}$Na} & \textbf{$^{39}$K} & \textbf{$^{24}$Mg} & \textbf{$^{48}$Ti} & \textbf{$^{57}$Ni} & \textbf{$^{59}$Co} & \textbf{$^{31}$P} & \textbf{$^{32}$S} \\ \hline

\hline
& \multicolumn{13}{c}{\textbf{EARTH}} \\ \hline

\hline
\textbf{Crust wt\%} & 46.7 & 27.7 & 8.1 & 5.1 & 3.7 & 2.8 & 2.6 & 2.1 & 0.6 & - & - & - & - \\

\hline
\textbf{Mantle wt\%} & 44.3 & 21.3 & 2.3 & 6.3 & 2.5 & - & - & 22.3 & - & 0.2 & - & - & - \\

\hline
\textbf{Core wt\%} & - & - & - & 84.5 & - & - & - & - & - & 5.6 & 0.3 & 0.6 & 9.0 \\
\hline

\hline
& \multicolumn{13}{c}{\textbf{MARS}} \\ \hline

\hline
\textbf{Mantle wt\%} & 39.2 & 16.2 & 1.2 & 23.7 & 1.4 & - & - & 18.3 & - & - & - & - & - \\

\hline
\textbf{Core wt\%} & - & - & - & 63.6 & - & - & - & - & - & - & - & - & 36.4 \\
\hline

\end{tabular}}
\end{center}
\end{table}

\begin{figure}[ht]
    \begin{center}
        \includegraphics[width=0.9\linewidth]{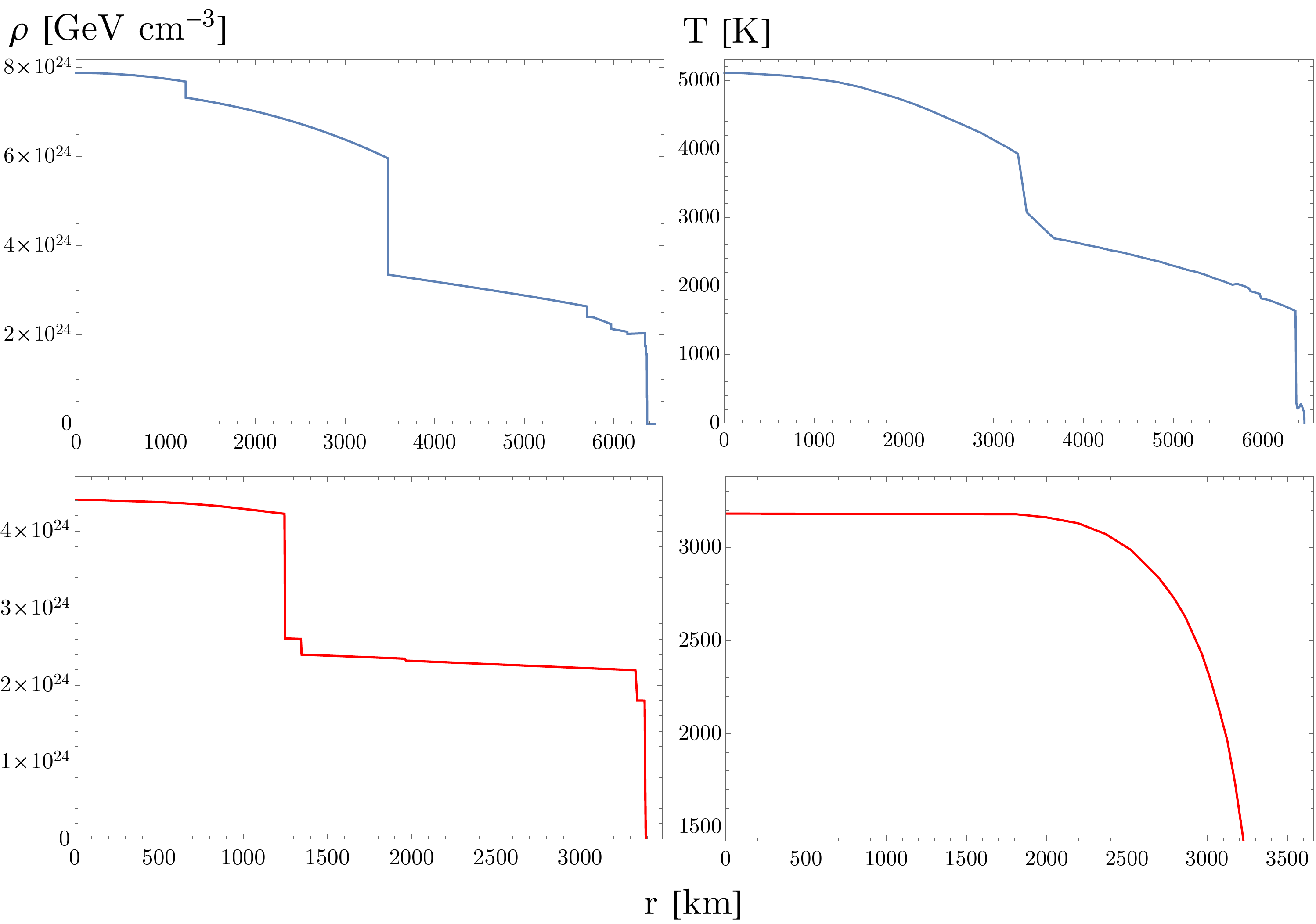}
        \captionsetup{justification=centering}
        \caption{The mass density (left) and temperature (right) profiles of the Earth (top) and Mars (bottom) used in this work.} 
        \label{fig:PREM}
    \end{center}
\end{figure}

The Earth's density also varies as a function of distance from the center. The preliminary reference Earth model \cite{DZIEWONSKI1981297} is a reasonable approximation of Earth's mass density. To be conservative, we used the minimum possible Martian density at all radii from the models given in \cite{johnston1974,JOHNSTON1977}. We henceforth reference these densities as $\rho(r)$, which is plotted in Figure \ref{fig:PREM}. To convert $\rho(r)$ to $n_j(r)$, the number density of element $j$, we divide by the mass of that element $m_j$ and multiply by the mass fraction of that element for either the core, mantle, or crust, as given in Table \ref{tab:SIcomp}.

It is also necessary to determine the planets' temperature profiles. There have been multiple Earth temperature models proposed; there is not strong consensus on the matter \cite{arevalo2009, earle2015, JAUPART2015223}. To be conservative, we take the highest reasonable proposed temperature at all radii to construct a "maximum temperature profile" of the Earth. This is shown in Figure \ref{fig:PREM}. The atmospheric temperatures used for Earth can be found in \cite{NASA1962}. We use the highest possible temperature profile in \cite{johnston1974} as our model for Mars, and used the Martian atmospheric temperatures given in \cite{biver2005}.

Many geological studies and models have been performed that estimate the total heat flux from within the Earth, all of which find the total heat flux to be around 44 TW. Some of this flux has been attributed to known processes, such as emission from radiogenic sources like uranium and thorium present in the Earth \cite{Buffett1675,JAUPART2015223}. However, to be conservative we will attribute all of the observed 44 TW to DM annihilation, when setting bounds on DM parameters. We similarly take a total heat flux of 3.5 TW for Mars, which is the maximum value of the projected range given in \cite{parro2017}.

To calculate spin-dependent DM cross sections, we must determine the spin properties of the various elements in the Earth. The natural abundances of several elements with non-zero nuclear spins ($J$) have been tabulated in \cite{deLaeter2009}. The values used in this work can be found in Table \ref{tab:SDcomp}, which gives the fraction of the corresponding element in Table \ref{tab:SIcomp} that exists as the non-zero nuclear spin isotope listed. 

\begin{table}[ht]
\begin{center}
\captionsetup{justification=centering}
\caption{Number percentages of elements of interest with non-zero nuclear spin in the crust, mantle, and core. }
\label{tab:SDcomp}
\resizebox{\textwidth}{!}{
\begin{tabular}{c c c c c c c c c c c c c c c c}\hline
& \textbf{$^{17}$O} & \textbf{$^{29}$Si} & \textbf{$^{27}$Al} & \textbf{$^{57}$Fe} & \textbf{$^{43}$Ca} & \textbf{$^{23}$Na} & \textbf{$^{39}$K} & \textbf{$^{25}$Mg} & \textbf{$^{47}$Ti} & \textbf{$^{49}$Ti} & \textbf{$^{61}$Ni} & \textbf{$^{31}$P} & \textbf{$^{33}$S} \\ \hline

\hline
\textbf{Number \%} & 0.4 & 4.7 & 100 & 2.12 & 0.135 & 100 & 100 & 10 & 7.44 & 5.41 & 1.14 & 100 & 0.75 \\

\hline
\end{tabular}}
\end{center}
\end{table}

Later references to number density, $n_j(r)$, will carry different meanings in the spin-independent and spin-dependent cases. For the former, it will refer to the number density of a given element. For the latter, it will refer to the number density of a given elemental isotope with non-zero nuclear spin.

In general, a nucleus is able to have spin if one or more of its nucleons is unpaired. Paired nucleons' spins will cancel, leading to a net-zero nuclear spin. However, even if all nucleons of a given kind are paired, the expectation value of the spin of paired protons or neutrons ($\ev{S_p}$ and $\ev{S_n}$ respectively) may be non-zero, which can result in spin-spin interactions. These terms are required to calculate spin-dependent per-nucleon cross sections (see Equation \ref{sigmaSD}), but their exact values remain unknown, and are slightly model dependent. The values used in this work are tabulated for the elements of interest in Table \ref{tab:spinParams}.

\begin{table}[ht]
\begin{center}
\captionsetup{justification=centering}
\caption{Spin parameters of elements of interest in the Earth and Mars \cite{Bednyakov2004}. }
\label{tab:spinParams}
\resizebox{\textwidth}{!}{%
\begin{tabular}{c c c c c c c c c c c c c c c}\hline
& \textbf{$^{17}$O} & \textbf{$^{29}$Si} & \textbf{$^{27}$Al} & \textbf{$^{57}$Fe} & \textbf{$^{43}$Ca} & \textbf{$^{23}$Na} & \textbf{$^{39}$K} & \textbf{$^{25}$Mg} & \textbf{$^{47}$Ti} & \textbf{$^{49}$Ti} & \textbf{$^{61}$Ni} & \textbf{$^{59}$Co} & \textbf{$^{31}$P} & \textbf{$^{33}$S} \\ \hline

\hline
$\boldsymbol{J}$ & 5/2 & 1/2 & 5/2 & 1/2 & 7/2 & 3/2 & 3/2 & 5/2 & 5/2 & 7/2 & 3/2 & 7/2 & 1/2 & 3/2 \\

\hline
$\boldsymbol{ \langle S_p \rangle}$ & -0.036 & 0.054 & 0.333 & 0 & 0 & 0.2477 & -0.196 & 0.04 & 0 & 0 & 0 & 0.5 & 0.181 & 0 \\

\hline
$\boldsymbol{ \langle S_n \rangle}$ & 0.508 & 0.204 & 0.043 & 1/2 & 1/2 & 0.0199 & 0.055 & 0.376 & 0.21 & 0.29 & -0.357 & 0 & 0.032 & -0.3\\

\hline
\end{tabular}}
\end{center}
\end{table}

For the purpose of this work, we have taken both the Earth and Mars to be modeled as perfect spheres with isotropic densities, temperatures, and compositions. We also ignore atmospheric scattering of DM, as the bulk of atoms within these planets is far larger and therefore dominates all scattering.

\section{Planetary Heating Limits on Annihilating Dark Matter}
\label{sec:lims}
With our planetary compositions and DM interaction models established, we will now bound DM's couplings using anomalous heating of the Earth and Mars. Results are given in terms of $m_\chi$, $\sigma_{\chi N}$, and the DM-DM annihilation cross section $\sigma_{\chi\bar{\chi}}$. Full exclusion limits obtained by requiring s-wave DM annihilations to not exceed the energy emitted from Earth's surface are shown in Figures \ref{fig:SIsimple}, \ref{fig:SDsimple}, \ref{fig:SDpsimple}, and \ref{fig:SDnsimple}. A full set of plots for s-wave, p-wave, and impeded \cite{Kopp:2016yji} dark matter annihilation for both Earth and Mars can be found in the appendix.

\subsection{Total Annihilation}\label{perfAnn}
To set a lower bound on the DM-nucleon cross section, we first consider the case that all captured DM annihilates, so that the rate at which DM is captured by the Earth or Mars is also the rate at which it annihilates. As we will see, this capture-annihilation equilibrium is reached in all parameter space of interest for a DM s-wave annihilation cross section $\sigma_{\chi \bar \chi} \gtrsim 10^{-34}~{\rm cm^2}$. 

To compute a limit for the case of DM capture-annihilation equilibrium, we began by running one thousand Monte Carlo simulations for each point on a grid of $m_\chi$-$\sigma_{\chi N}$ values. 
We chose to run one thousand simulations, for the following reasons. The total potential dark matter energy flux through the Earth, as described by equation \ref{completeHeat}, is around 3300 TW. Because our exclusion condition requires 44 TW of heating, only $44/3300 \approx 0.01$ of the velocity normalized dark matter flux must be captured to set a bound. By running a thousand simulations, we expect to have sampled $\sim 10$ times more of the DM distribution than is strictly necessary to set a bound. For each of the thousand simulations, the DM particle in question was given a random initial velocity, distributed by a three dimensional Maxwell-Boltzmann distribution 
\begin{equation}
\label{dPdv}
    f(v) = \frac{v_i^3}{N_{e}} \exp{\left(-\frac{3 \vec v_d^2}{2\sigma_{v}^2}\right)},
\end{equation}

\noindent where this expression is the rate-normalized Maxwell-Boltzmann distribution \cite{Smith:1988kw}. We took standard values of $v_0=220$ km s$^{-1}$, with $\sigma_v=v_0\sqrt{3/2}$ as the velocity dispersion, and with the velocity of DM in the planet's rest frame defined as $\vec v_d = \vec v_i+ \vec v_e $ where $|\vec v_e| \approx 230$ km s$^{-1}$ is the Earth's velocity in the galactic rest frame \cite{Smith:1988kw,Bramante:2016rdh}. We normalize $N_e$ in this Maxwell Boltzmann distribution to match a conservatively low background DM density of $\rho_\chi = 0.3~{\rm GeV/cm^3}$ \cite{Iocco:2011jz,Buch:2018qdr,lisanti2007} and truncate the distribution at a conservatively low galactic escape velocity of $v_{esc}=528$ km s$^{-1}$ \cite{Deason2019}. 

For each simulated DM particle, an entry angle $\theta$ into the planet in question was randomly chosen, distributed according to the probability density function

\begin{equation}
\label{dPdt}
\frac{dP(\theta)}{d\theta}=2\sin(\theta)\cos(\theta),
\end{equation}

\noindent which yields the average chord length traveled through the Earth, as described by Dirac's formula \cite{SJOSTRAND20021607}. Using the generated $\theta$ and $v_i$ values in Equations \ref{eq:vf}, \ref{taubar}, and \ref{eq:si} or \ref{sigmaSD} for the spin-independent or spin-dependent cases respectively, and using Poisson-distributed $\tau_j$ (see Equation \ref{poisson}), we found $v_f$ for all angles generated.

\begin{figure}[t!]
    \centering
    \includegraphics[width=0.87\textwidth]{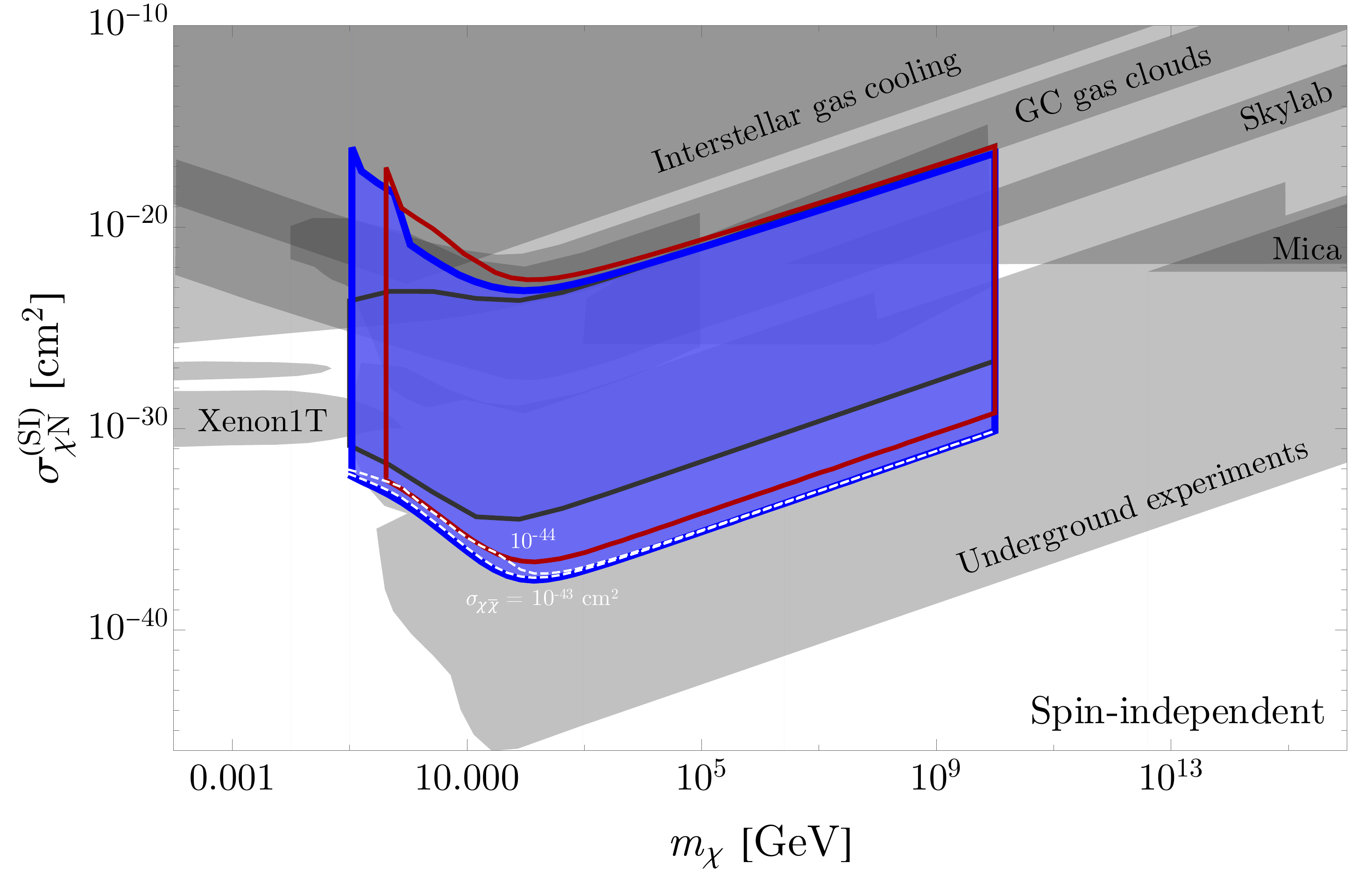}
    \caption{Spin-independent DM-nucleon Earth heating limit assuming all DM annihilates (blue) and for DM self-annihilation cross sections $\sigma_{\chi\bar{\chi}}$ given in cm$^2$ as labelled, and the prospective Mars heating exclusion limit for $\sigma_{\chi\bar{\chi}}=10^{-36}$ cm$^2$ (red). The edges of these exclusion limits are determined as follows: the bottom edge is set by requiring a large enough cross section that enough DM is captured and anomalously heats the planet, the top edge is set by requiring that, despite having a large scattering cross section, DM can drift to the center of the planet within a billion years, and the left edge requires that DM is not so light that it evaporates out of the planet's interior. While the right edge is fixed at $10^{10}$ GeV by s-wave annihilation unitarity considerations, the bounds shown could be consistently extended to higher masses. Juxtaposed are limits from \cite{Mack2007,XENON1T,Overburden,XQC,Skylab,CMB1,CMB2,GasClouds1, GasClouds2,mica1,mica2,mica3}. Underlaid, the dark grey line shows the Earth heating bound set by Mack et al.~\cite{Mack2007}. A number of recent references \cite{mica2,Digman:2019wdm} have pointed out that only a few dark matter models will consistently provide an effective DM-nucleon cross section in excess of $\sigma_{\chi N} \approx 10^{-26}~{\rm cm^2}$. For a brief discussion of dark matter models which validly imply a DM-nucleon cross section greater than $\sigma_{\chi N} \approx 10^{-26}~{\rm cm^2}$, see the last paragraph of Section \ref{sec:cap}.}
    \label{fig:SIsimple}
\end{figure}

In each simulation, we found $P_{cap}$, the probability of capturing a single DM particle entering the Earth, for a given DM mass and cross section. Defining $\ev{v_i}$ as the average initial velocity of a captured DM particle, the mass capture rate ($\Gamma_{m}$) was calculated using the following equation for the total mass capture rate

\begin{equation}
\label{completeHeat}
    \Gamma_{m} = 2 \pi R_e^2 \ev{v_i} \rho_\chi P_{cap},
\end{equation}
where we note that this is half the total flux of DM expected through the Earth's surface, since we are only interested in ingoing (and not outgoing) DM particles.

\begin{figure}[t!]
    \centering
    \includegraphics[width=0.87\textwidth]{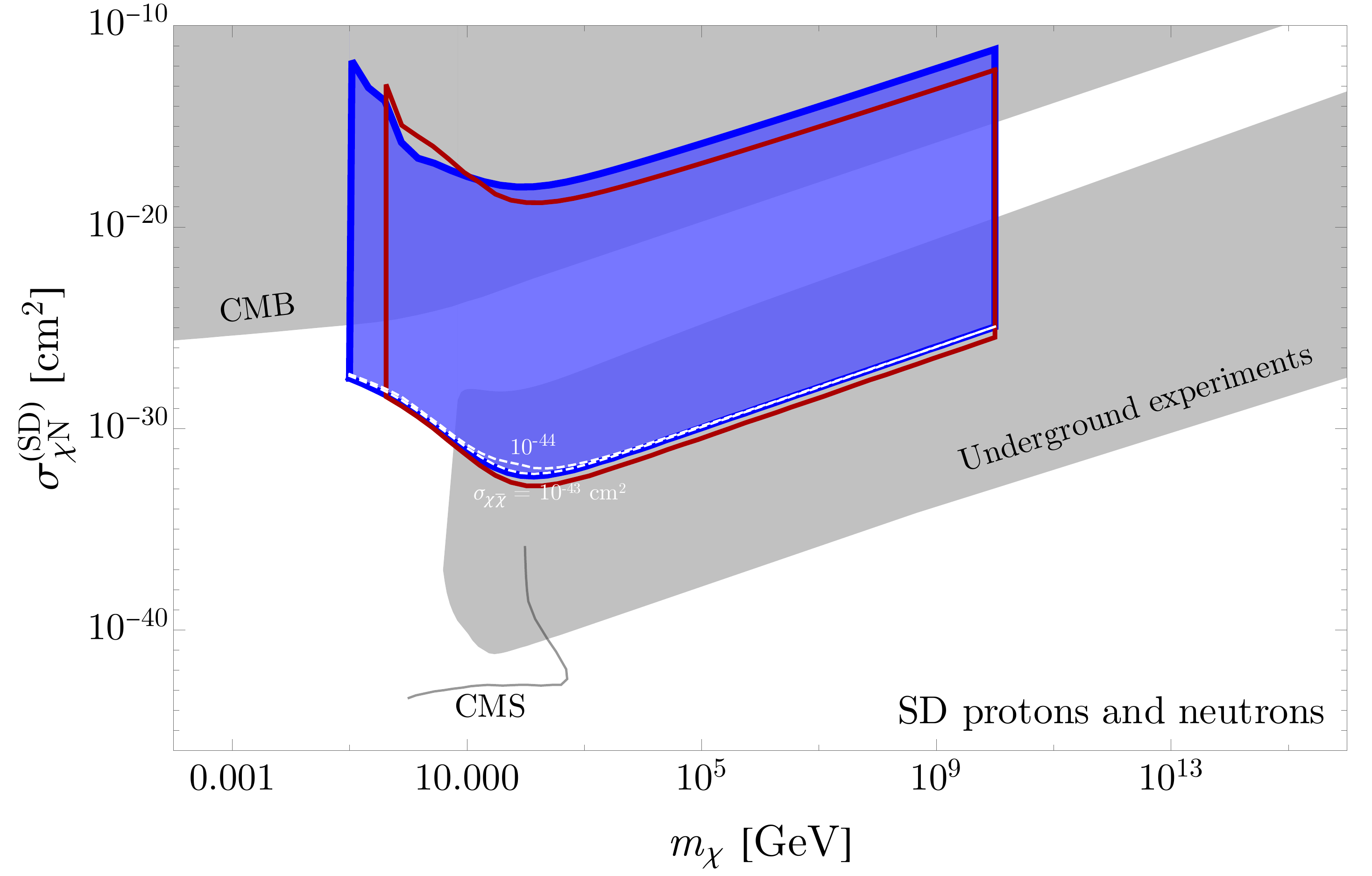}
    \caption{Spin-dependent DM-nucleon Earth heating limit (blue) for DM self-annihilation cross sections $\sigma_{\chi\bar{\chi}}$ as labelled, and the Mars heating exclusion limit for $\sigma_{\chi\bar{\chi}}=10^{-36}$ cm$^2$ (red). Juxtaposed with limits given by \cite{PICO1,PICO2,CMS,XENON1T,CMB2}. This exclusion assumes equal spin-dependent coupling to neutrons and protons, aka isospin-independent scattering with $a_n=a_p=1$ in Eq.~\eqref{sigmaSD}.}
    \label{fig:SDsimple}
\end{figure}

\begin{figure}[h!]
    \centering
    \includegraphics[width=0.87\textwidth]{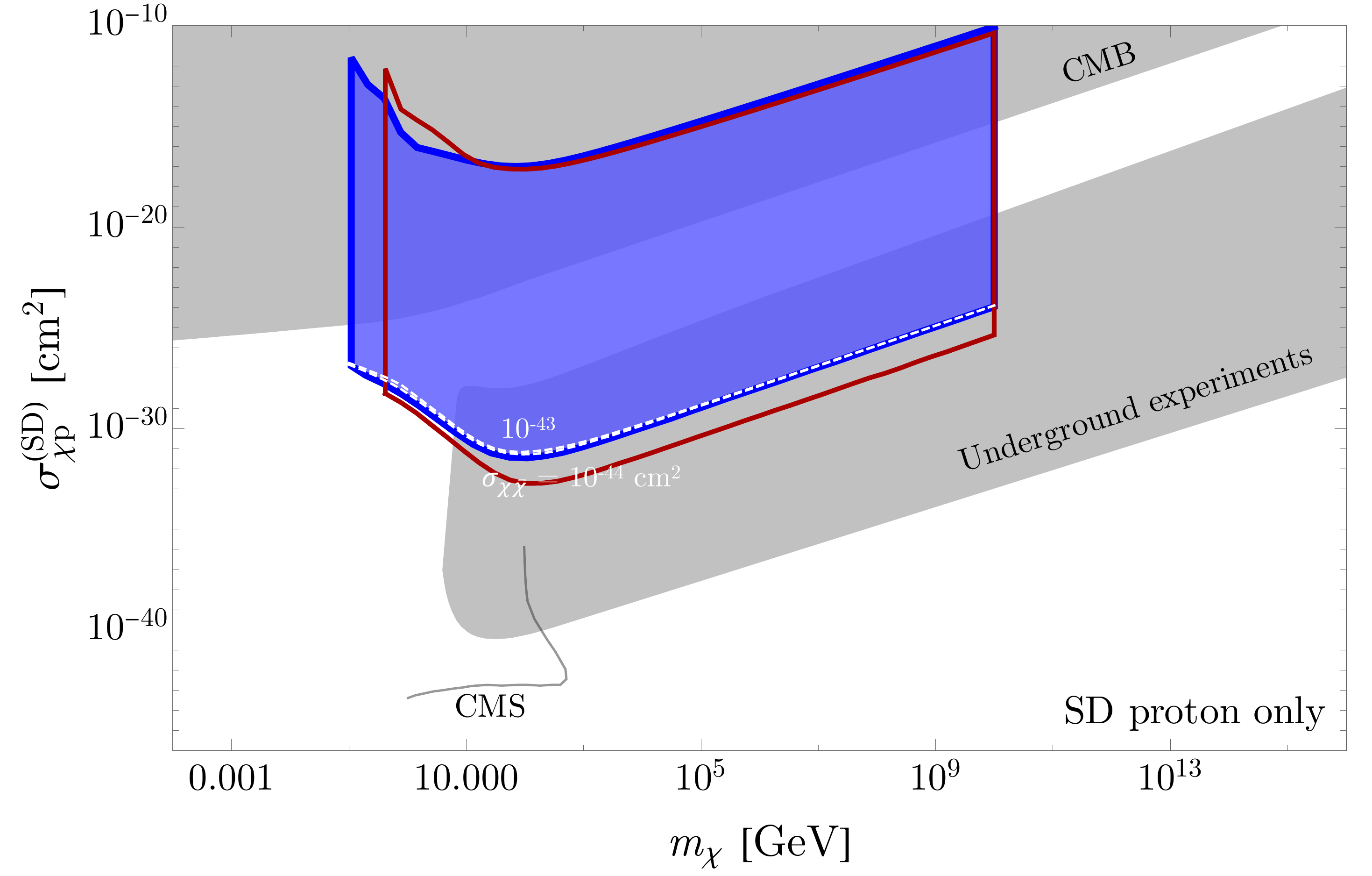}
    \caption{Spin-dependent DM-proton Earth heating limit (blue) for DM self-annihilation cross sections $\sigma_{\chi\bar{\chi}}$ as labelled, and the Mars heating exclusion limit for $\sigma_{\chi\bar{\chi}}=10^{-36}$ cm$^2$ (red). Juxtaposed with limits given by \cite{PICO1,PICO2,CMB2}. This exclusion assumes only spin-dependent coupling to protons, with $a_p=1$ and $a_n=0$ in Eq.~\eqref{sigmaSD}.}
    \label{fig:SDpsimple}
\end{figure}

\begin{figure}[h!]
    \centering
    \includegraphics[width=0.87\textwidth]{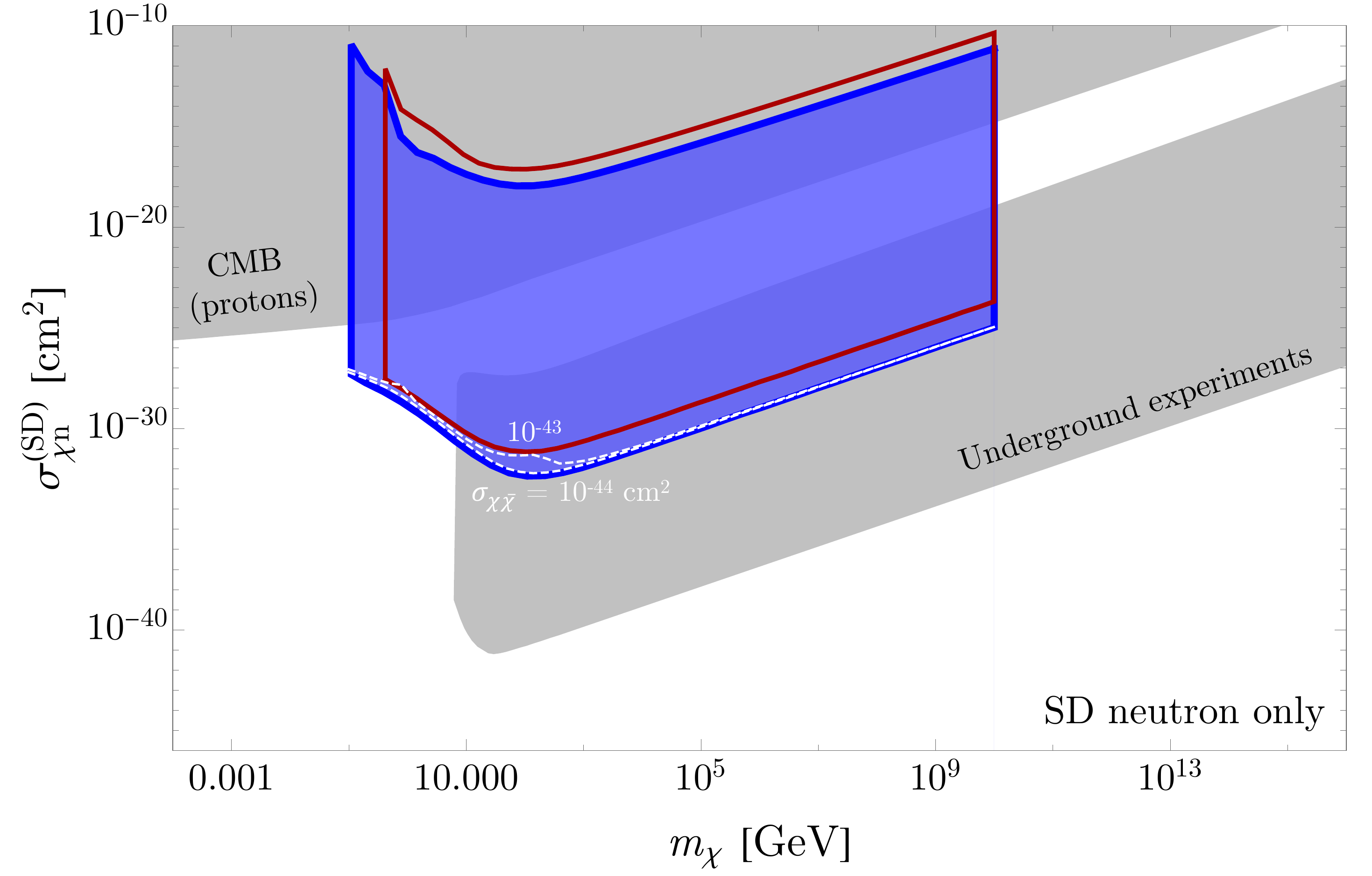}
    \caption{Spin-dependent DM-neutron Earth heating limit (blue) for DM self-annihilation cross sections $\sigma_{\chi\bar{\chi}}$ as labelled, and the Mars heating exclusion limit for $\sigma_{\chi\bar{\chi}}=10^{-36}$ cm$^2$ (red). Juxtaposed with limits given by \cite{XENON1T,CMB2}. This exclusion assumes only spin-dependent coupling to neutrons, with $a_p=0$ and $a_n=1$ in Eq.~\eqref{sigmaSD}.}
    \label{fig:SDnsimple}
\end{figure}

By simulating one thousand DM particles per parameter space point tested, we identified the minimum nucleon scattering cross section for which more than $\sim$44 TW of DM would be captured by the Earth. We continued simulation iterations, until the range of cross section values converged upon varied by less than one percent, and among these values selected a cross section that implied slightly more than 44 TW of heating by DM. Assuming that all captured DM annihilates, this amount of darkogenic Earth heating can be safely excluded, since the total heat output of the Earth is 44 TW. This method was then repeated for Mars, this time using a maximum heat flux of 3.5 TW. The masses and cross sections that resulted in $\dot{M}_\chi \approx44$ TW in the Earth are shown as the lower solid blue limits in Figures \ref{fig:SIsimple}, \ref{fig:SDsimple}, \ref{fig:SDpsimple}, and \ref{fig:SDnsimple}. The values generated for Mars are shown as solid red limits. The limit has been truncated at $m_\chi=10^{10}$ GeV, a mass cutoff advocated in Ref.~\cite{Mack2007}, chosen by requiring that the non-relativistic DM s-wave self-annihilation cross section required for DM to reach capture-annihilation equilibrium not exceed a unitarity limit \cite{Griest1990}. The planetary heating bounds in this paper can be trivially extended to higher masses, though for such large masses and self-annihilation cross sections, it would be preferable to consider an explicit model.

\subsection{Thermal Evaporation}\label{evap}
A gravitationally bound DM particle will not necessarily stay captured. Thermally vibrating nuclei in the Earth or Mars may scatter with captured DM and increase the DM's kinetic energy. For a sufficiently low DM mass, this kinetic energy increase can result in DM escaping its planet's gravity. This process is known as evaporation (see $e.g.$ \cite{Bramante2014}). Evaporation implies a minimum mass for which Earth heating exclusions are valid, since evaporated DM will not annihilate within the planet.

First we consider the thermal radius, which is the radius of containment for DM that has reached thermal equilibrium with the planet - where DM's average thermal energy will be equal to its average potential energy in the Earth's gravitational well. The thermal radius is found using the virial theorem

\begin{equation}
\label{rth}
   \frac{1}{2}m_\chi \left(v_{th}^{(\chi)}(r_{th})\right)^2=\frac{3}{2}T(r_{th})=\frac{GM(r_{th})m_\chi}{2r_{th}},
\end{equation}

DM masses less than $\sim$100 MeV yield thermal radii greater than the radius of the Earth, meaning that the sphere of containment for DM at this mass extends outside the Earth. For Mars, this mass is $\sim$420 MeV. To be conservative, we end our limit at this mass, as the upper exclusion limit loses its meaning when there is no defined thermal radius to which a DM particle would drift. 

A caveat to the evaporation process as detailed above is that, for moderately large cross sections, low mass particles can be re-captured as they scatter against terrestrial or Martian constituent particles along their exit trajectory. Using the root mean square thermal velocity of a particle in equilibrium with the Earth or Mars, $
    v_{th}(r)=\sqrt{\frac{3T(r)}{m}},
$ one could determine for what cross section light DM would be re-captured, and the new effective volume within which DM annihilates.  We discuss what future work might be done along these lines in Section \ref{sec:conc}.

\subsection{Partial Annihilation}\label{impann}
For a small enough self-annihilation cross section, DM that is efficiently captured by the Earth or Mars may not lead to anomalous heating. Including the effect of DM annihilation, the differential equation describing the number of DM particles accumulated is

\begin{equation}
\label{dNdt}
    \frac{dN_\chi}{dt}\approx C_\chi-\frac{N_\chi^2\ev{\sigma_{\chi\bar{\chi}}v_{th}^{(\chi)}}}{V_{th}},
\end{equation}

\noindent where $C_\chi=\dot{M}_\chi/m_\chi$ is the number of DM particles captured per second and $V_{th}=4\pi r_{th}^3/3$ is the thermalized volume. Note that $V_{th}$ will decrease with increasing $m_\chi$. As is customary in the literature, we define the s-wave, p-wave, and impeded \cite{Kopp:2016yji} dark matter annihilation rates such that the thermal velocity dependence of the DM annihilation rate in Eq.~\eqref{dNdt} scales as \mbox{ $\ev{\sigma_{\chi\bar{\chi}}v_{th}^{(\chi)}} = \left\{ \sigma_{\chi\bar{\chi}},~\sigma_{\chi\bar{\chi}}v_{th}^2,~\sigma_{\chi\bar{\chi}}v_{th} \right\}$}, respectively. Solving Equation \ref{dNdt} gives an expression for the number of DM particles captured over a period of time. For convenience, we define
$
    c_{ann}=\ev{\sigma_{\chi\bar{\chi}}v_{th}^{(\chi)}}/V_{th},
$
which gives a compact expression for the number of DM particles in the planet, accounting for self-annihilation $N_\chi$,
\begin{equation}
\label{N}
    N_\chi=\sqrt{\frac{C_\chi}{c_{ann}}}\tanh\left(\sqrt{C_\chi c_{ann}}\cdot t\right).
\end{equation}
DM will have reached capture-annihilation equilibrium once $t \sim 1 / \sqrt{C_\chi c_{ann}}$.  

The equation for the annihilation rate of DM is then
$
    C_{ann} = \frac{c_{ann}N_\chi^2}{2},
$
and each DM-DM annihilation will have an energy of approximately $2m_\chi$. Thus, we define a new equation for the heating rate induced by DM $\dot{Q}_\chi$ that accounts for the DM-DM annihilation cross section \cite{GREEN2019120}:

\begin{equation}
\label{heat}
    \dot{Q}_\chi=2m_\chi C_{ann}=m_\chi C_{\chi}\tanh^2\left(\sqrt{C_\chi c_{ann}}\cdot t\right).
\end{equation}
Again simulating many DM capture events as described in Section \ref{perfAnn}, but now also using the self-annihilation cross section to determine the total heating rate as described above, we determined values of $m_\chi$, $\sigma_{\chi\bar{\chi}}$, and $\sigma_{\chi N}$ corresponding to a $C_{\chi}$ that yielded a heat flux of $\sim$44 TW. By running simulations with multiple combinations of these three parameters, we found exclusion limits for values of $\sigma_{\chi\bar{\chi}}$, shown as dashed lines in Figures \ref{fig:SIsimple}, \ref{fig:SDsimple}, \ref{fig:SDpsimple}, \ref{fig:SDnsimple}. Note that for a sufficiently large DM-DM s-wave annihilation cross section, the white partial annihilation lines converge on the total annihilation limit, at $\sigma_{\chi \bar{\chi}} \gtrsim 10^{-44}$ cm$^2$ for both Earth Mars.

\subsection{Drift Time}\label{driftTime}
DM with a large enough nuclear cross section will be captured efficiently by the Earth and Mars. However, too large a nuclear cross section will result in DM stopping near the surface of the planet. As a result, it would be contained in a larger volume than DM that is free to descend to its thermalization radius. With its annihilation rate volumetrically suppressed, the DM will not necessarily annihilate quickly enough to heat the planet in which it is captured. Here we set an upper cross section limit by ensuring that the captured DM is able to drift through the planet in question on a timescale less than the planet's age.

To compute this upper limit on our cross section exclusion, we used a  treatment similar to \cite{Mack2007, starkman1990, GOULD1990337}. Assuming DM-nuclear interactions are frequent enough, the planet's gravitational force will balance against viscous drag. We require DM to drift from the surface of the planet to its thermal radius (described in section \ref{evap}) within 4.5 Gyr, the approximate age of both the Earth and Mars. Balancing these two forces yields

\begin{equation}
\label{gravDrag1}
    \frac{GM(r)m_\chi}{r^2}=v_{drift} \left\{\sum_{j}n_j(r)m_j\ev{\sigma_{\chi j} v_j(r)}\right\},
\end{equation}

\noindent where $G$ is the gravitational constant, $v_j(r)=\sqrt{3
T(r)/m_j}$ is the thermal velocity of molecule $j$, $M(r)$ is the mass enclosed in radius $r$ (equal to the volume integral over $\rho(r)$), and $v_{drift}=\partial r / \partial t$ is the drift velocity of the DM \cite{GOULD1990337}. Making these substitutions, Equation \ref{gravDrag1} becomes

\begin{equation}
\label{gravDrag2}
    t=\frac{1}{Gm_\chi}\sum_{j}\left\{\sigma_{\chi j} \int_{r_{th}}^{Re} dr \cdot r^2n_j(r) \sqrt{3m_jT(r)}\left[4\pi\int_{0}^r dr'\cdot r'^2\rho(r')\right]^{-1}\right\}
\end{equation}

\noindent where we conservatively set t $=$ 1 Gyr in calculations. Our upper exclusion limits, shown as the upper solid lines in Figures \ref{fig:SIsimple}, \ref{fig:SDsimple}, \ref{fig:SDpsimple}, \ref{fig:SDnsimple}, were found by solving Equation \ref{gravDrag2} for $\sigma_{\chi N}$ at a given $m_\chi$. In the spin-independent case, we used Equation \ref{eq:si} to convert from per-nucleon to nuclear cross section. In the spin-dependent case, we used Equation \ref{sigmaSD}, and values given in Table \ref{tab:spinParams}.

\section{Discussion}\label{sec:conc}
We have derived Earth and Mars heating bounds on both spin-dependent and spin-independent DM scattering. Of course, these bounds will apply to DM which annihilates to Standard Model particles. DM that does not annihilate, or which annihilates to particles which freely stream out of the Earth or Mars are not excluded by this result. We have also determined how Earth and Mars heating bounds change as DM's self-annihilation cross section is varied.

The limits we have found on dark matter annihilation in the Earth are especially interesting, considering that the canonical WIMP self-annihilation cross section is $\sigma_{\chi\bar{\chi}} \sim10^{-36}$ cm$^2$ \cite{JUNGMAN1996195}. While the Earth heating limit appears to exclude the canonical WIMP self-annihilation cross section for s-wave dark matter annihilation, p-wave and impeded annihilating dark matter are not entirely excluded. As such, a major result presented in this work for the first time, is that DM planetary heating bounds do not necessarily constrain p-wave and impeded DM models.

In future work, there are additional improvements that could be made to the treatment of DM heating planets. For lighter DM, heating is limited by dark matter evaporation. The treatment of DM evaporation presented here can be improved using Monte Carlo simulations that account for incoming light DM scattering repeatedly not only during capture, but also during evaporation. Captured DM that interacts strongly enough with nuclei, may avoid evaporation by being trapped via back-scattering. Such a future analysis might result in better limits on light DM. 

The potential effectiveness of this analysis for other planets in our solar system should also be considered. While the Earth and Martian interior compositions are known with most certainty, based on seismic data and surface mineral sampling, relatively thermally inactive bodies such as Earth's moon, which has a maximum DM capture rate of $\sim$250 TW, and a projected internal heat flow of $\sim$0.75 TW \cite{moonflow}, could be studied to set additional limits on DM. Analysis of lunar chemical composition indicates a significant presence of heavier elements with non-zero nuclear spin \cite{moonComp}, meaning that improved spin-dependent DM scattering limits may be possible. On the other hand, temperatures very near the Moon's center are still largely unknown, and would require further study.

Finally, for the case of DM with predominantly spin-dependent nucleon interactions, our results have shown that substantial parameter space exists for low dark matter masses, where experiments like PICO might be able to set new limits on or discover strongly-interacting DM by placing detectors above-ground. We leave a detailed study of the prospects for above-ground spin-dependent dark matter searches to future work.

\section*{Acknowledgements}
We thank Javier Acevedo and Rebecca Leane for useful conversations, and especially Bill McDonough, Ondrej Sramek, Laura Sammon, Yufei Xi, Scott Wipperfurth, and Hiroko Watanabe for their geological expertise. 

\appendix

\section{Full Exclusion Limits}
Below we detail Earth heat flow bounds, and prospective Mars bounds, for s-wave, p-wave, and impeded DM self-annihilation cross sections, $\sigma_{\chi\bar{\chi}}$, as labelled. For ease of comparison and consistency we have truncated all bounds at $m_\chi =10^{10}~{\rm GeV}$, which is an s-wave unitarity limit on the equilibration annihilation cross section as first advocated in \cite{Mack2007}. Subject to DM model building considerations, all bounds could be consistently extended to higher masses.
 \newpage
\subsection{S-Wave Annihilation}

\begin{figure}[!htbp]
    \centering
    \includegraphics[width=0.9\textwidth]{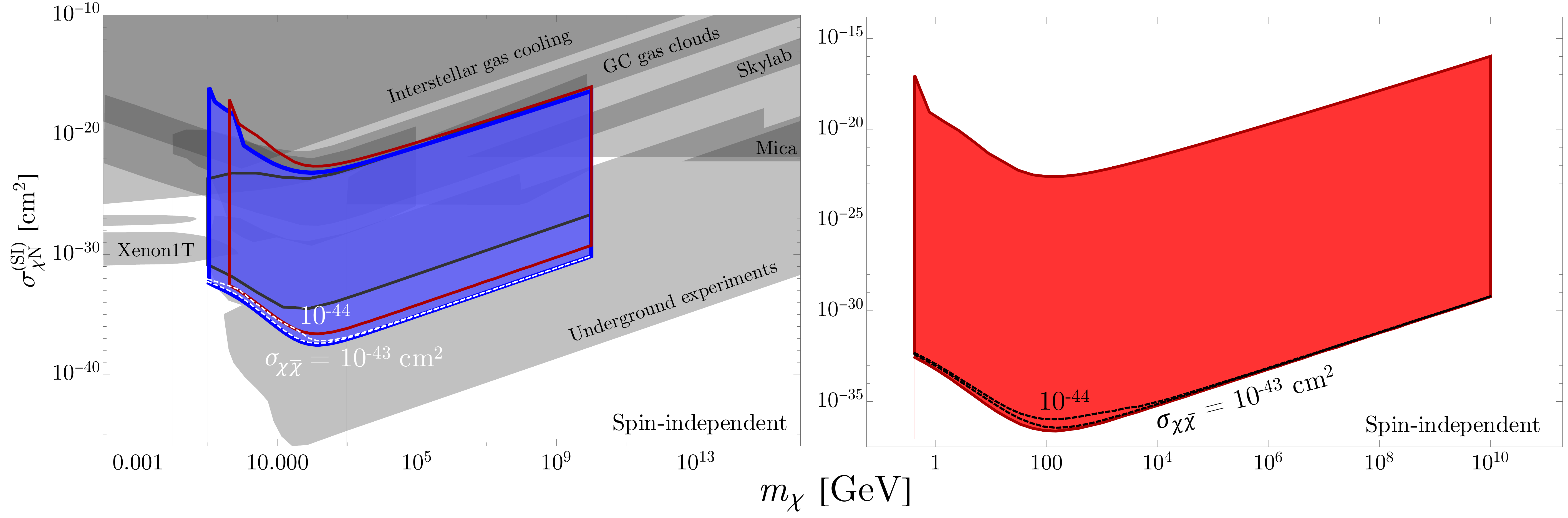}
    \caption{Spin-independent DM-nucleon Earth (left) and prospective Mars (right) heating exclusion limits for DM s-wave self-annihilation cross sections $\sigma_{\chi\bar{\chi}}$ as labelled.}
    \label{fig:SI_Swave_Duo}
\end{figure}

\begin{figure}[!htbp]
    \centering
    \includegraphics[width=0.9\textwidth]{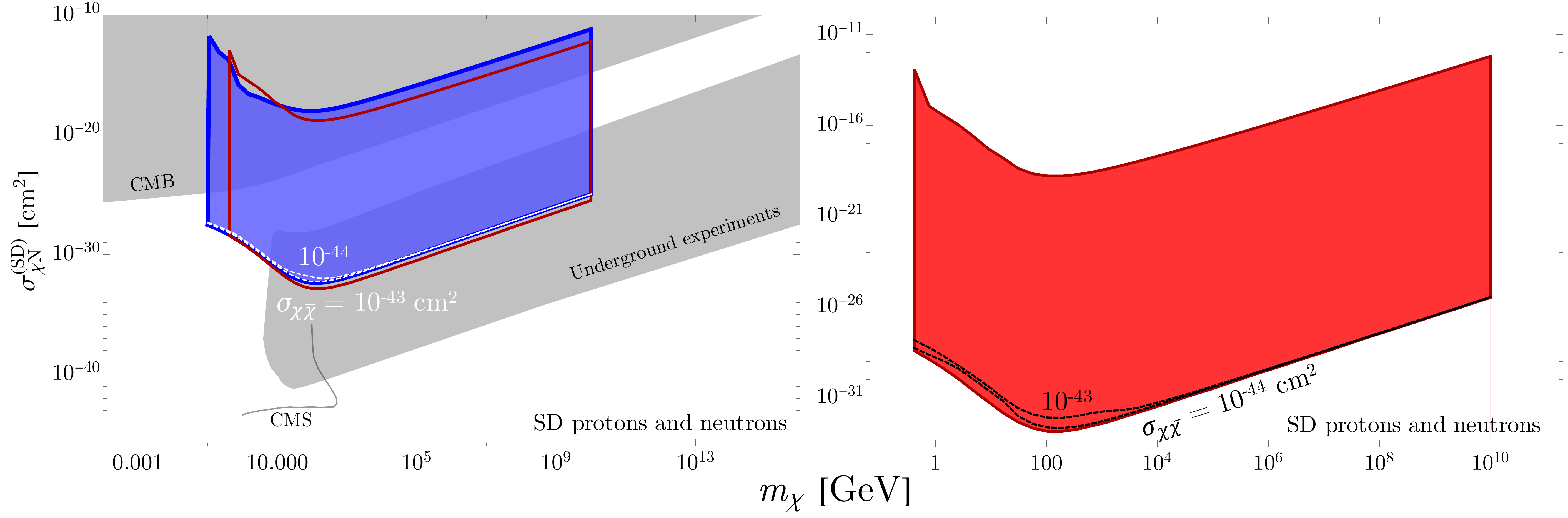}
    \caption{Spin-dependent DM-nucleon Earth (left) and prospective Mars (right) heating exclusion limits for DM s-wave self-annihilation cross sections $\sigma_{\chi\bar{\chi}}$ as labelled, assuming equal spin-dependent coupling to neutrons and protons, aka isospin-independent scattering with $a_n=a_p=1$ in Eq.~\eqref{sigmaSD}.}
    \label{fig:SD_Swave_Duo}
\end{figure}

\begin{figure}[!htbp]
    \centering
    \includegraphics[width=0.9\textwidth]{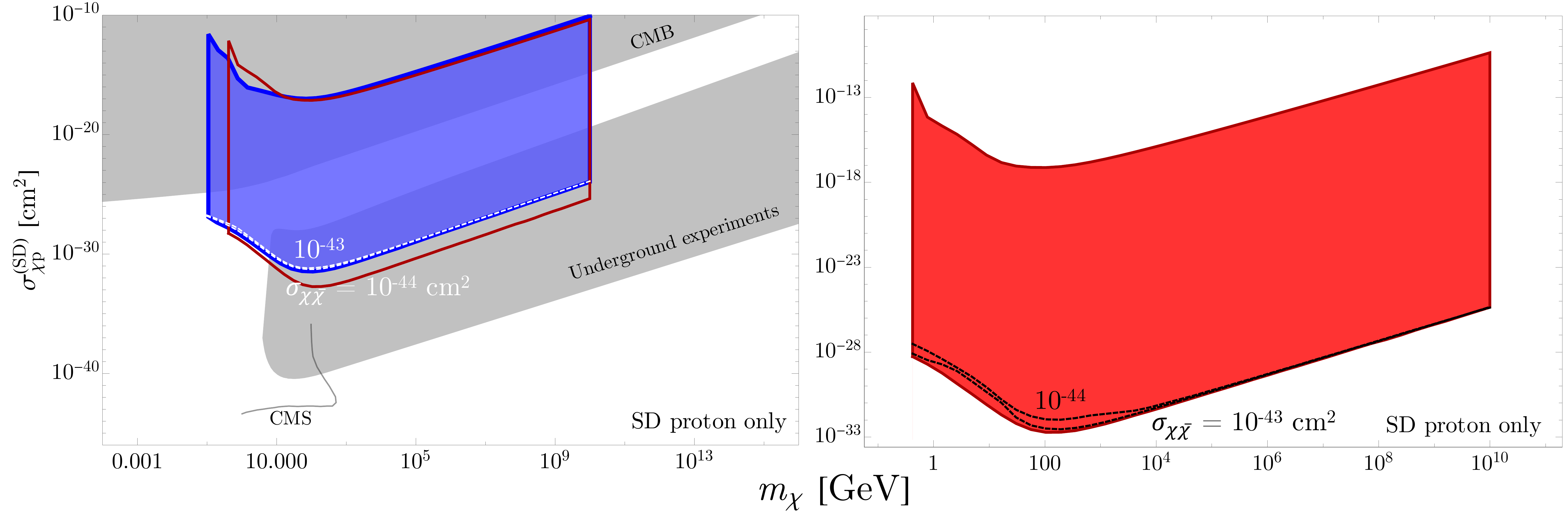}
    \caption{Spin-dependent DM-nucleon Earth (left) and prospective Mars (right) heating exclusion limits for DM s-wave self-annihilation cross sections $\sigma_{\chi\bar{\chi}}$ as labelled, assuming spin-dependent coupling to protons, with $a_p=1$ and $a_n=0$ in Eq.~\eqref{sigmaSD}.}
    \label{fig:SDp_Swave_Duo}
\end{figure}

\begin{figure}[!htbp]
    \centering
    \includegraphics[width=0.9\textwidth]{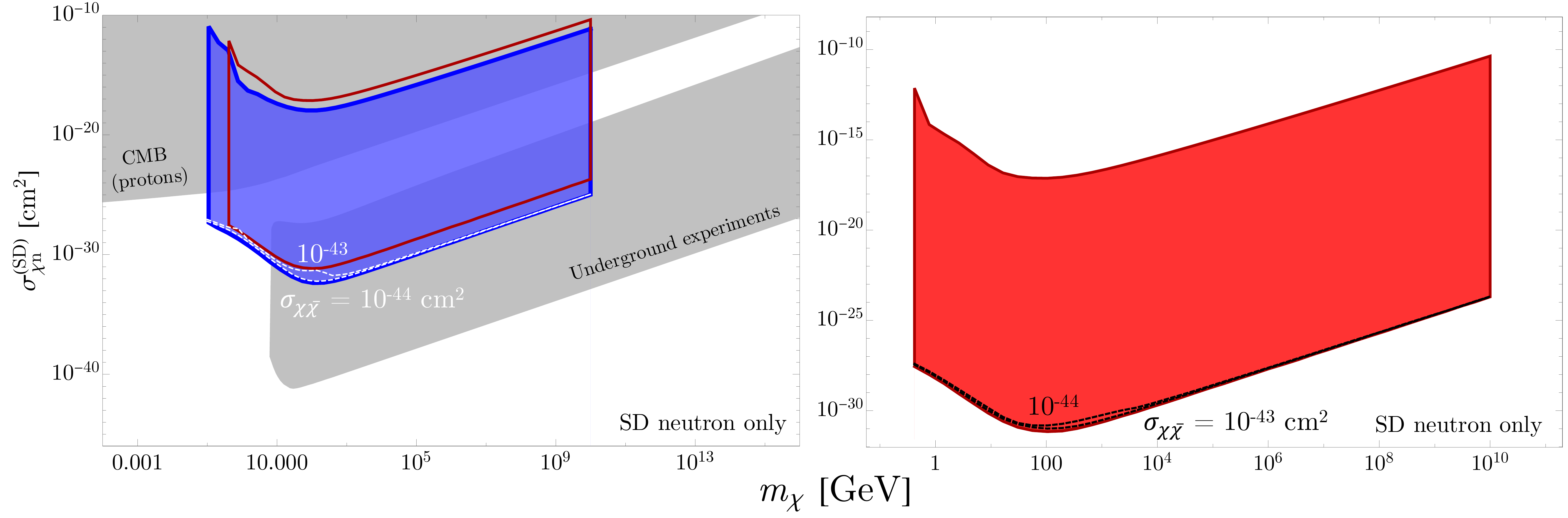}
    \caption{Spin-dependent DM-nucleon Earth (left) and prospective Mars (right) heating exclusion limits for DM s-wave self-annihilation cross sections $\sigma_{\chi\bar{\chi}}$ as labelled, assuming spin-dependent coupling to neutrons, with $a_p=0$ and $a_n=1$ in Eq.~\eqref{sigmaSD}.}
    \label{fig:SDn_Swave_Duo}
\end{figure}
\FloatBarrier
\subsection{P-Wave Annihilation}
\begin{figure}[!htbp]
    \centering
    \includegraphics[width=0.9\textwidth]{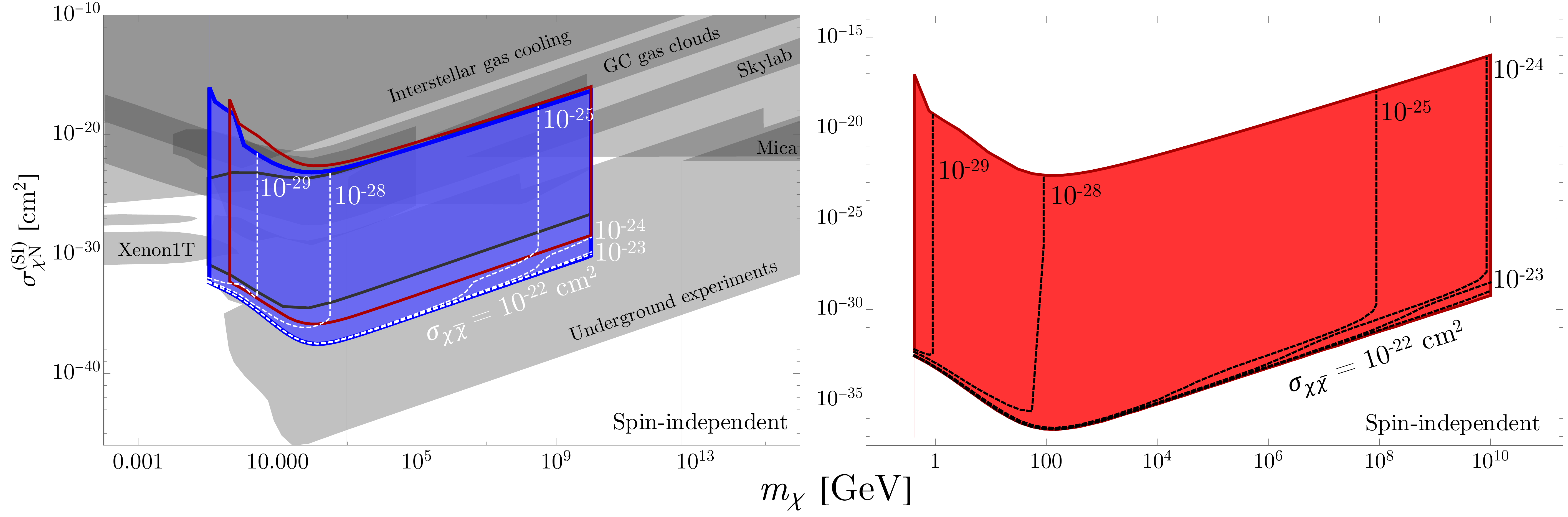}
    \caption{Spin-independent DM-nucleon Earth (left) and prospective Mars (right) heating exclusion limits for DM p-wave self-annihilation cross sections $\sigma_{\chi\bar{\chi}}$ as labelled.}
    \label{fig:SI_Pwave_Duo}
\end{figure}

\begin{figure}[!htbp]
    \centering
    \includegraphics[width=0.9\textwidth]{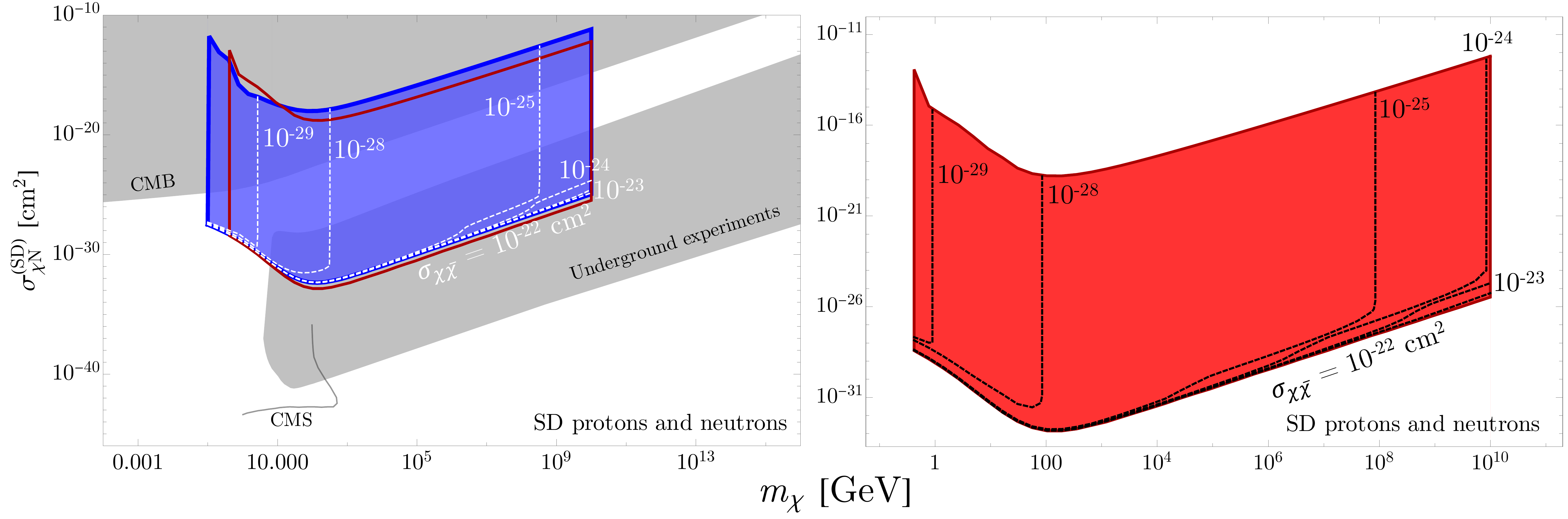}
    \caption{Spin-dependent DM-nucleon Earth (left) and prospective Mars (right) heating exclusion limits for DM p-wave self-annihilation cross sections $\sigma_{\chi\bar{\chi}}$ as labelled, assuming equal spin-dependent coupling to neutrons and protons, aka isospin-independent scattering with $a_n=a_p=1$ in Eq.~\eqref{sigmaSD}.}
    \label{fig:SD_Pwave_Duo}
\end{figure}

\begin{figure}[!htbp]
    \centering
    \includegraphics[width=0.9\textwidth]{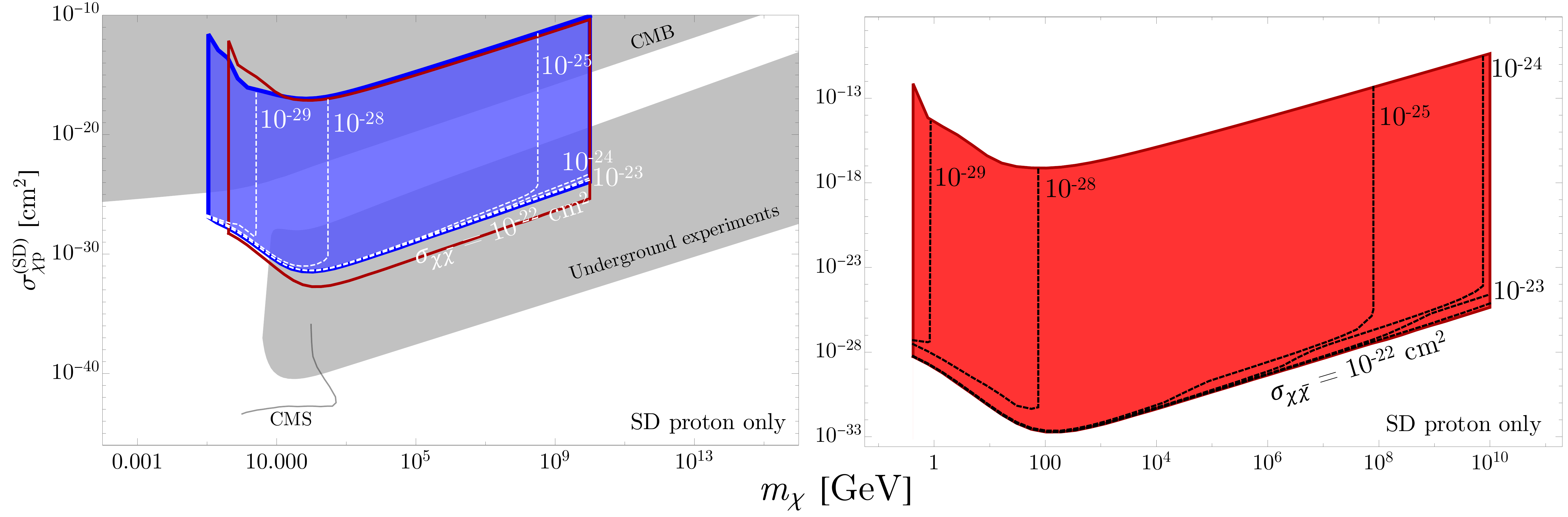}
    \caption{Spin-dependent DM-nucleon Earth (left) and prospective Mars (right) heating exclusion limits for DM p-wave self-annihilation cross sections $\sigma_{\chi\bar{\chi}}$ as labelled, assuming spin-dependent coupling to protons, with $a_p=1$ and $a_n=0$ in Eq.~\eqref{sigmaSD}.}
    \label{fig:SDp_Pwave_Duo}
\end{figure}

\begin{figure}[H]
    \centering
    \includegraphics[width=0.9\textwidth]{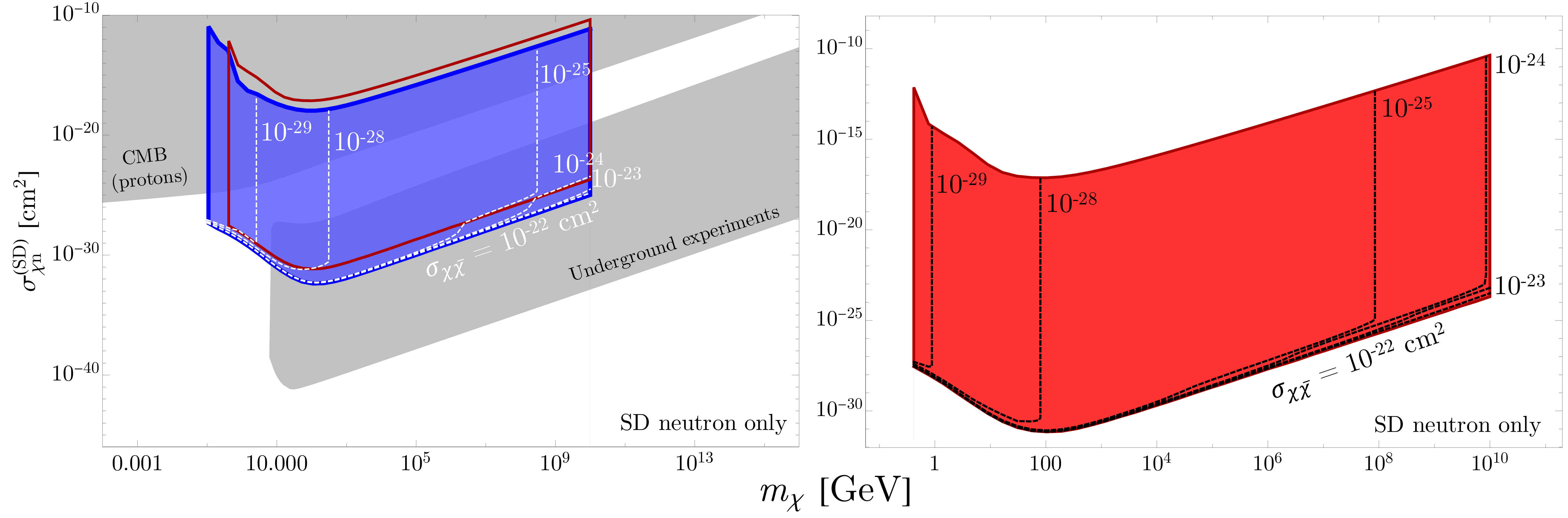}
    \caption{Spin-dependent DM-nucleon Earth (left) and prospective Mars (right) heating exclusion limits for DM p-wave self-annihilation cross sections $\sigma_{\chi\bar{\chi}}$ as labelled, assuming spin-dependent coupling to neutrons, with $a_p=0$ and $a_n=1$ in Eq.~\eqref{sigmaSD}.}
    \label{fig:SDn_Swave_Duo}
\end{figure}
\FloatBarrier
\subsection{Impeded Annihilation}
\begin{figure}[H]
    \centering
    \includegraphics[width=0.9\textwidth]{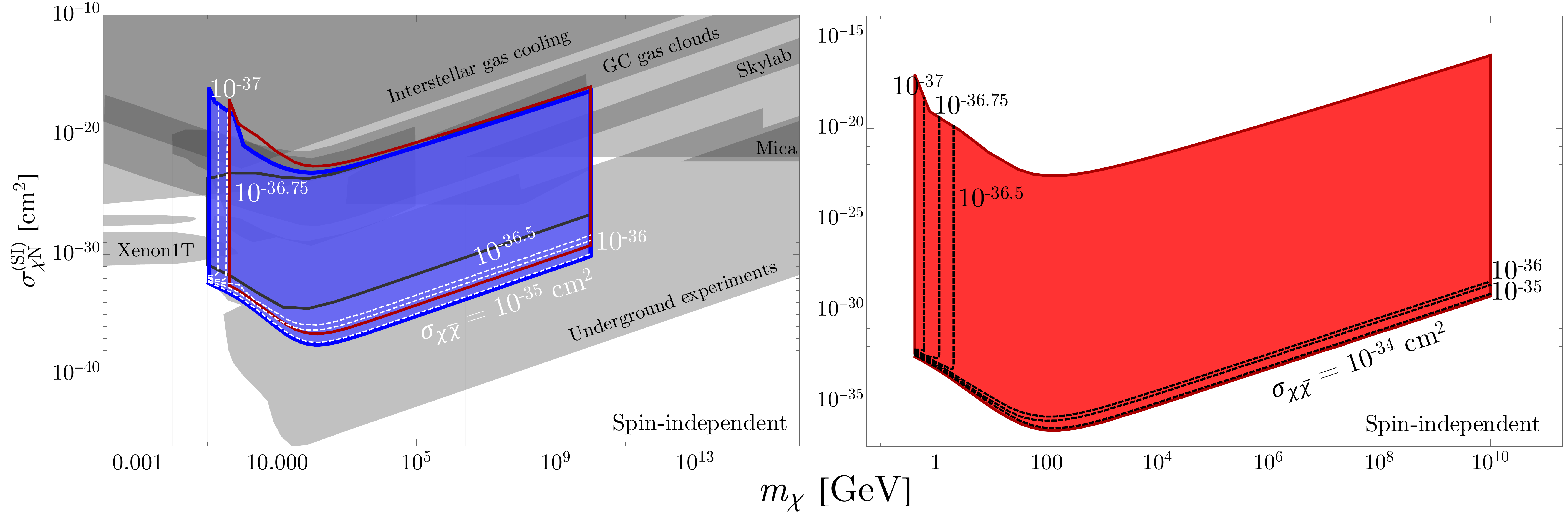}
    \caption{Spin-independent DM-nucleon Earth (left) and prospective Mars (right) heating exclusion limits for DM impeded self-annihilation cross sections $\sigma_{\chi\bar{\chi}}$ as labelled.}
    \label{fig:SI_Imp_Duo}
\end{figure}

\begin{figure}[H]
    \centering
    \includegraphics[width=0.9\textwidth]{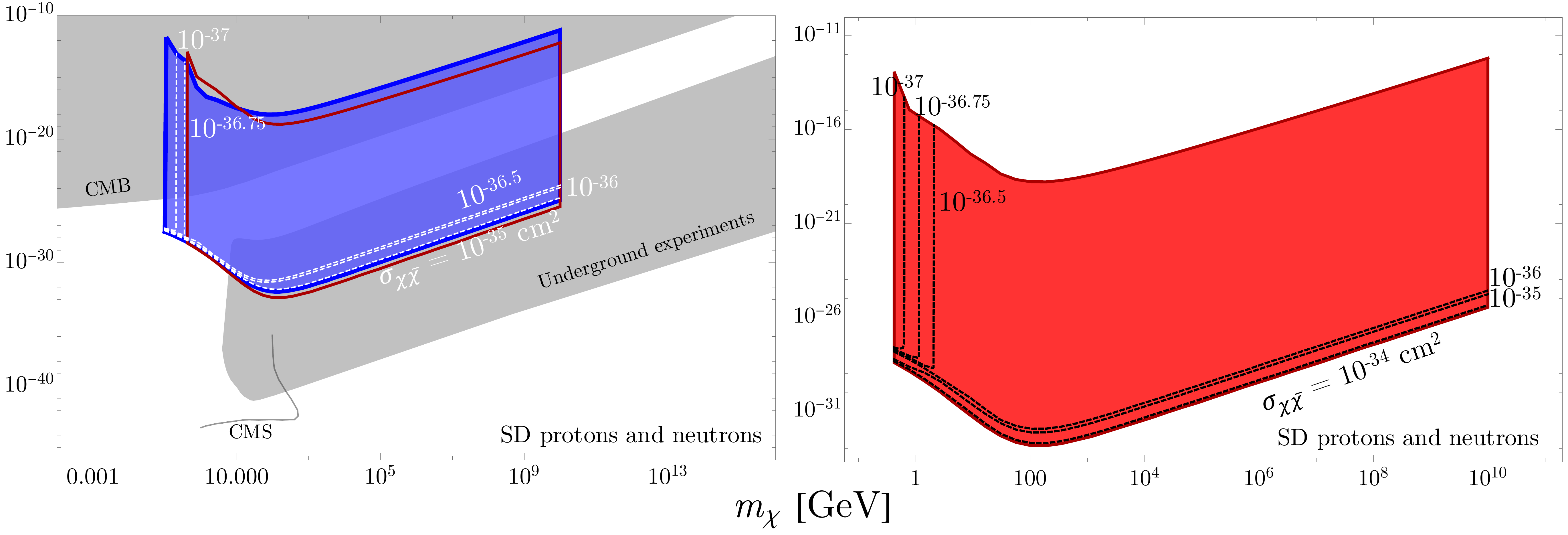}
    \caption{Spin-dependent DM-nucleon Earth (left) and prospective Mars (right) heating exclusion limits for DM impeded self-annihilation cross sections $\sigma_{\chi\bar{\chi}}$ as labelled, assuming equal spin-dependent coupling to neutrons and protons, aka isospin-independent scattering with $a_n=a_p=1$ in Eq.~\eqref{sigmaSD}.}
    \label{fig:SD_Imp_Duo}
\end{figure}

\begin{figure}[H]
    \centering
    \includegraphics[width=0.9\textwidth]{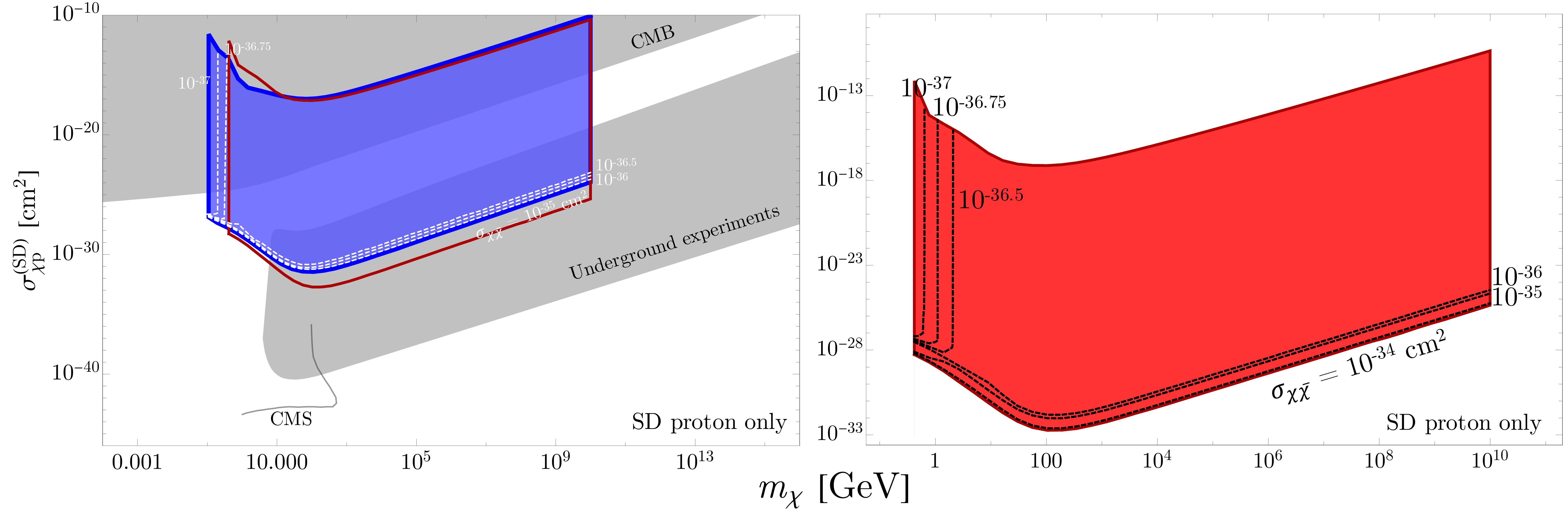}
    \caption{Spin-dependent DM-nucleon Earth (left) and prospective Mars (right) heating exclusion limits for DM impeded self-annihilation cross sections $\sigma_{\chi\bar{\chi}}$ as labelled, assuming spin-dependent coupling to protons, with $a_p=1$ and $a_n=0$ in Eq.~\eqref{sigmaSD}.}
    \label{fig:SDp_Imp_Duo}
\end{figure}

\begin{figure}[H]
    \centering
    \includegraphics[width=0.9\textwidth]{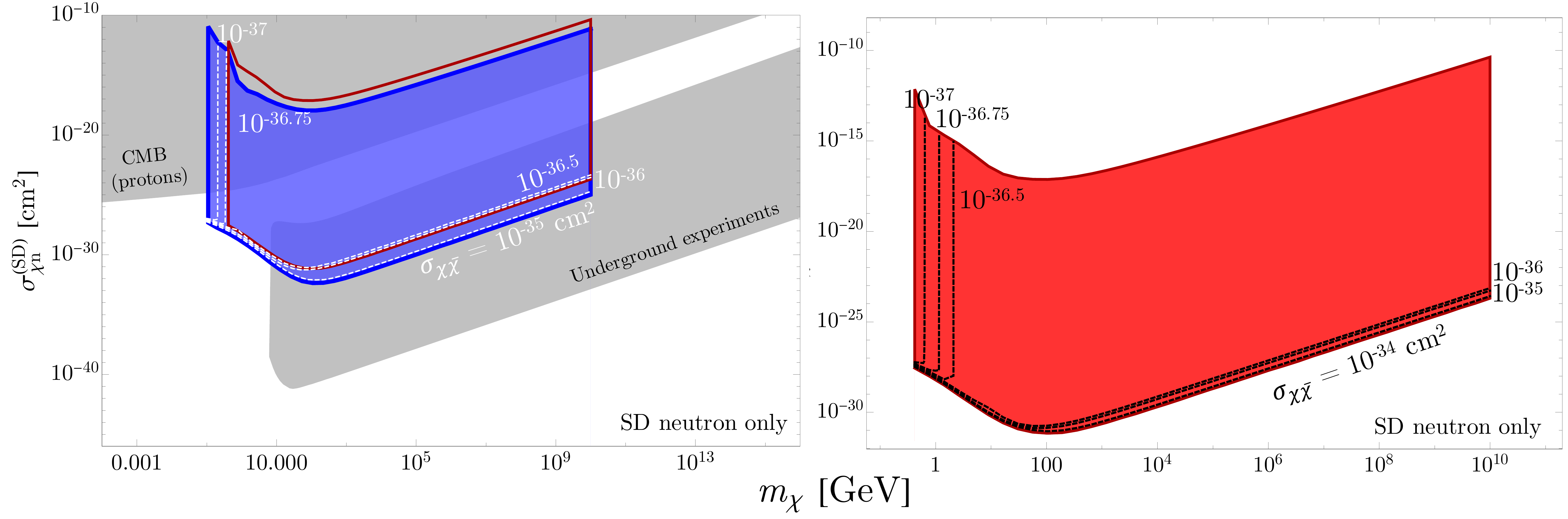}
    \caption{Spin-dependent DM-nucleon Earth (left) and prospective Mars (right) heating exclusion limits for DM impeded self-annihilation cross sections $\sigma_{\chi\bar{\chi}}$ as labelled, assuming spin-dependent coupling to neutrons, with $a_p=0$ and $a_n=1$ in Eq.~\eqref{sigmaSD}.}
    \label{fig:SDn_Imp_Duo}
\end{figure}

\newpage
 
\bibliography{EM}
\bibliographystyle{JHEP}

\end{document}